\documentclass[proof]{pasj01}

\Received{$\langle$reception date$\rangle$}
\Accepted{$\langle$acception date$\rangle$}
\Published{$\langle$publication date$\rangle$}
\usepackage[T1]{fontenc}
\usepackage{color}

\begin{document}

\title{Development of a new wideband heterodyne receiver system for the Osaka 1.85-m mm-submm telescope\large{\\--- Corrugated horn \& Optics covering 210--375\,GHz band ---}}
\author{Yasumasa \textsc{Yamasaki}\altaffilmark{1}
}

\altaffiltext{1}{Department of Physical Science, Graduate School of Science, Osaka Prefecture University, 1-1 Gakuen-cho, Naka-ku, Sakai, Osaka 599-8531, Japan}

\altaffiltext{2}{Institute of Astronomy, Graduate School of Science, University of Tokyo, 2-21-1 Osawa, Mitaka, Tokyo 181-0015, Japan}

\altaffiltext{3}{National Astronomical Observatory of Japan, Advanced Technology Center, 2-21-1 Osawa, Mitaka, Tokyo 181-8588, Japan}

\altaffiltext{4}{The Graduate University for Advanced Studies, SOKENDAI, 2-21-1 Osawa, Mitaka, Tokyo 181-8588, Japan}

\email{s\_y.yamasaki@p.s.osakafu-u.ac.jp}
\author{Sho \textsc{Masui}\altaffilmark{1}}
\author{Hideo \textsc{Ogawa}\altaffilmark{1}}
\author{Hiroshi \textsc{Kondo}\altaffilmark{1}}
\author{Takeru \textsc{Matsumoto}\altaffilmark{1}}
\author{Masanari \textsc{Okawa}\altaffilmark{1}}
\author{Koki \textsc{Yokoyama}\altaffilmark{1}}
\author{Taisei \textsc{Minami}\altaffilmark{1}}
\author{Ryotaro \textsc{Konishi}\altaffilmark{1}}
\author{Sana \textsc{Kawashita}\altaffilmark{1}}
\author{Ayu \textsc{Konishi}\altaffilmark{1}}
\author{Yuka \textsc{Nakao}\altaffilmark{1}}
\author{Shimpei \textsc{Nishimoto}\altaffilmark{1}}
\author{Sho \textsc{Yoneyama}\altaffilmark{1}}
\author{Shota \textsc{Ueda}\altaffilmark{1}}
\author{Yutaka \textsc{Hasegawa}\altaffilmark{1}}
\author{Shinji \textsc{Fujita}\altaffilmark{1}}
\author{Atsushi \textsc{Nishimura}\altaffilmark{1, 2}}
\author{Takafumi \textsc{Kojima}\altaffilmark{3, 4}}
\author{Keiko \textsc{Kaneko}\altaffilmark{3}}
\author{Ryo \textsc{Sakai}\altaffilmark{3}}
\author{Alvaro \textsc{Gonzalez}\altaffilmark{3, 4}}
\author{Yoshinori \textsc{Uzawa}\altaffilmark{3, 4}}
\author{Toshikazu \textsc{Onishi}\altaffilmark{1}}

\maketitle

\begin{abstract}
The corrugated horn is a high performance feed often used in radio telescopes. There has been a growing demand for wideband optics and corrugated horns in millimeter and submillimeter-wave receivers.  It improves the observation efficiency and allows us to observe important emission lines such as CO in multiple excited states simultaneously. However, in the millimeter/submillimeter band, it has been challenging to create a conical corrugated horn with a fractional bandwidth of $\sim$60\% because the wavelength is very short, making it difficult to make narrow corrugations. In this study, we designed a conical corrugated horn with good return loss, low cross-polarization, and symmetric beam pattern in the 210--375\,GHz band (56\% fractional bandwidth) by optimizing the dimensions of the corrugations. The corrugated horn was installed on the Osaka 1.85-m mm-submm telescope with the matched frequency-independent optics, and simultaneous observations of \textrm{$^{12}$}CO, \textrm{$^{13}$}CO, and C\textrm{$^{18}$}O ($J$ = 2--1, 3--2) were successfully made. In this paper, we describe the new design of the corrugated horn and report the performance evaluation results including the optics.
\end{abstract}

\section{Introduction}
We have been conducting survey observations of \textrm{$^{12}$}CO, \textrm{$^{13}$}CO, and C\textrm{$^{18}$}O ($J$ = 2--1) in the Galactic plane using the Osaka 1.85-m mm-submm telescope at Nobeyama Radio Observatory (\cite{Onishi2013}; Nishimura et al.\ \yearcite{Nishimura2015}, \yearcite{Nishimura2020}). As a new project, we are planning to extend the frequency band of the receiver to make maps of \textrm{$^{12}$}CO, \textrm{$^{13}$}CO, and C\textrm{$^{18}$}O ($J$ = 3--2) at the same time in addition to the previous observation lines. Our goal is to get closer to the understanding of the star formation process by statistically studying the physical properties of molecular clouds (kinetic temperature, hydrogen density, etc.). For this purpose, we are developing a wideband receiver that covers the frequency coverage of 210--375\,GHz (56\% of the fractional bandwidth), where the above emission lines can be obtained simultaneously.  The overall receiver development is described in \citet{Masui2021}.  This paper focuses on the horn and the optics and presents the performance when installed on the telescope.\par
A typical optics in radio astronomy is the Cassegrain-type reflector antenna, and it is important to illuminate the sub-reflector with an appropriate edge taper, low cross-polarization, and low sidelobe level to achieve high aperture efficiency. To meet this request, we often use a corrugated horn as a feed (\cite{Olver1994}). The corrugated horn produces a hybrid mode HE$_{11}$ in which 98\% of the radiation pattern is a fundamental Gaussian beam (\cite{Teniente2003}).  It has a low sidelobe level and low cross-polarization intensity, making it a suitable feed for millimeter and submillimeter-wave band optics. However, the corrugated horn has at least a few corrugations within a wavelength in the direction along the axis of symmetry with a depth of about 1/2 to 1/4 wavelength or sometimes deeper, and it is thus difficult to manufacture the corrugated horn particularly in the millimeter and submillimeter band.  It has been manufactured by electroforming or machining using an L-shaped tool. If the corrugation width is smaller and the aspect ratio, the depth over the width, is larger, machining becomes more difficult. \citet{Kimura2008} achieved high performance in the development in the 150\,GHz band by cutting the throat section separately and press-fitting the product into the remaining part.  This method is applied to the ALMA Band 4 receivers (\cite{Asayama2014}).  Corrugated horns with wider frequency bands are desired for the next-generation radio telescopes, and it is also important to accurately reproduce and fabricate the horns for broadband applications.\par
Corrugated horns are generally classified into a conical corrugated horn
 (\cite{Zhang1993}), which has a tapered structure, and a profiled corrugated horn (\cite{Olver1988}), which has a curved basic structure and normally has a more compact structure than the conical corrugated horn. Profiled corrugated horns have a more compact structure and a higher degree of freedom than the conical corrugated horn, so that the profiled corrugated horn can be extended to a wider frequency range with intensive optimization. For example, in \citet{Lee2020}, the design and measurement of a profiled corrugated horn of 275--500\,GHz band (fractional bandwidth of $\sim$60\%) were reported. The higher degree of freedom indicates the high complexity of the design. In addition, the frequency dependence does not follow clear patterns. It is a matter of the particular design, and therefore we need to be very careful in considering lots of frequencies in the design. On the other hand, the behavior is more "monotonic" in a conical corrugated horn.\par
The optics characteristics of the conical corrugated horn, such as the beam size and beam curvature radius, can be easily determined from the basic parameters; the flare angle of the horn and aperture size.  Therefore, it is possible to control the Gaussian beam emitted by the conical corrugated horn regardless of the frequency. In the case of the profiled horn, a simulation is required for matching with the optics.  Moreover, the manufacturing process of the conical corrugated horn is simple and time-saving, requiring only a cutting tool to hollow out the metal block and an L-shaped tool to cut the corrugations.  The recent numerical simulation shows that wideband conical corrugated horn or profiled corrugated horn with a fractional bandwidth of 50\% or more can be realized by changing depths of conical corrugated horn or profile (\cite{Gonzalez2017}). A conical corrugated horn designed according to this idea has also been reported in the 67--116 GHz band (\cite{Yagoubov2020}).\par
With this simplicity of the design and fabrication, we adopt the conical corrugated horn for the broadband receiver system covering 210--375\,GHz.  In section 2, the frequency-independent optics of the 1.85-m telescope is presented. In section 3, the detailed design of the corrugated horn is described.  In section 4, the measurement result of the performance is given, and the commissioning observations on the telescopes are shown in section 5.

\section{Optics}
The optics of the 1.85-m radio telescope is a beam propagation system based on the Nasmyth-Cassegrain focus (\cite{Onishi2013}). It consists of a main-reflector with a diameter of 1.85\,m, a sub-reflector, 2 plane mirrors, and 2 ellipsoidal mirrors (figure \ref{fig:telescope}). In order to cover the new frequency band, we updated the optics as shown below, and the physical parameters of the new optics are shown in table \ref{tab:optics}.\par

\begin{figure}[h]
 \begin{center}
  \includegraphics[width=12cm]{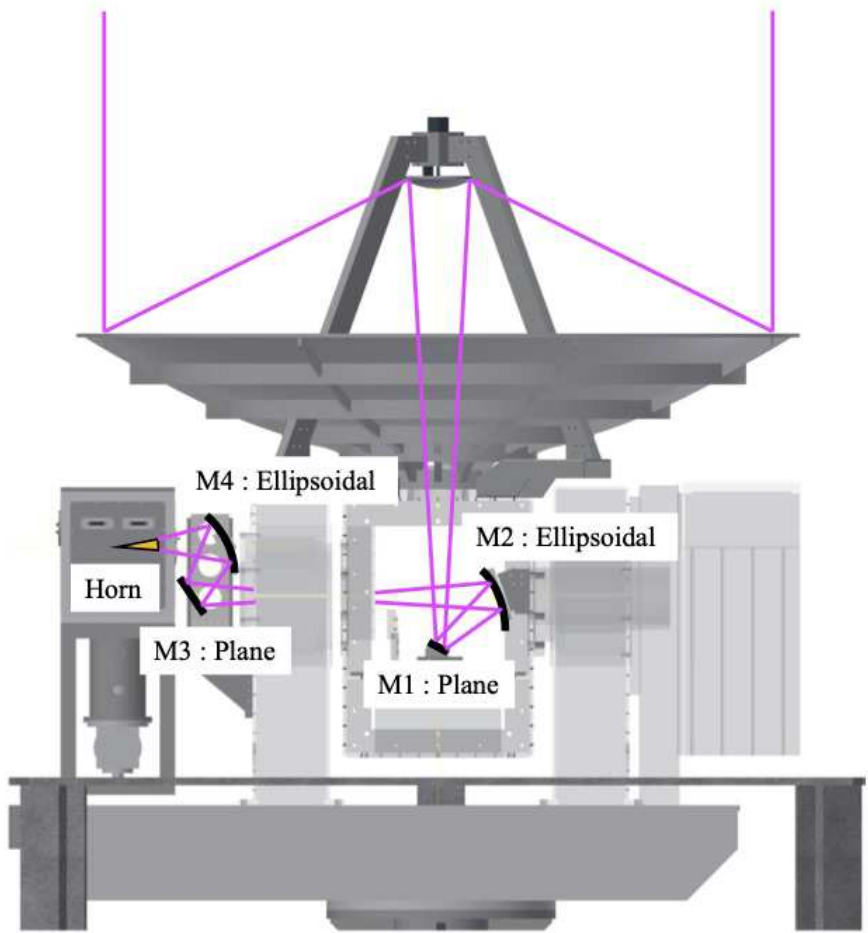} 
 \end{center}
\caption{CAD drawing of the 1.85-m mm-submm telescope optics. The telescope is enclosed in a radome to protect it from rain, wind, and sunlight, and to minimize temperature changes that can degrade the performance.}\label{fig:telescope}
\end{figure}

\begin{table}[h]
\caption{Physical parameters of the optics} 
\label{tab:optics}
\begin{center}       
\begin{tabular}{l l} 
\hline
\rule[-1ex]{0pt}{3.5ex}  Parameter & Value (mm)  \\
\hline\hline
\rule[-1ex]{0pt}{3.5ex}  Parameter of Main/Sub-ref.
 &   \\
\rule[-1ex]{0pt}{3.5ex}  \ \ \ \ Focal length of Main-ref.
 & 740.00  \\
\rule[-1ex]{0pt}{3.5ex}  \ \ \ \ Diameter of Sub-ref.
 & 185.00  \\
\rule[-1ex]{0pt}{3.5ex}  \ \ \ \ Interfocal distance of primary focus\\   \ \ \ \ \ and secondary focus
 & 1180.00  \\
\rule[-1ex]{0pt}{3.5ex}  Distance
 &   \\
\rule[-1ex]{0pt}{3.5ex}  \ \ \ \ Sub-ref. to M1(plane)
 & 1197.83  \\
\rule[-1ex]{0pt}{3.5ex}  \ \ \ \ M1 to M2 (ellipsoidal)
 & 202.17  \\
\rule[-1ex]{0pt}{3.5ex}  \ \ \ \ M2 to M3 (plane)
 & 825.00  \\
\rule[-1ex]{0pt}{3.5ex}  \ \ \ \ M3 to M4 (ellipsoidal)
 & 150.00  \\
\rule[-1ex]{0pt}{3.5ex}  \ \ \ \ M4 to horn aperture
 & 160.00  \\
\rule[-1ex]{0pt}{3.5ex}  Focal length of the ellipsoidal mirrors	
 &   \\
\rule[-1ex]{0pt}{3.5ex}  \ \ \ \ M2
 & 192.28  \\
\rule[-1ex]{0pt}{3.5ex}  \ \ \ \ M4
 & 131.93  \\
\rule[-1ex]{0pt}{3.5ex}  Parameters of horn	
 &   \\
\rule[-1ex]{0pt}{3.5ex}  \ \ \ \ Diameter of horn aperture
 & 8.00  \\
\rule[-1ex]{0pt}{3.5ex}  \ \ \ \ Slant length of horn
 & 32.49  \\
 
\hline
\end{tabular}
\end{center}
\end{table} 

\begin{figure}[h]
 \begin{center}
  \includegraphics[width=12cm]{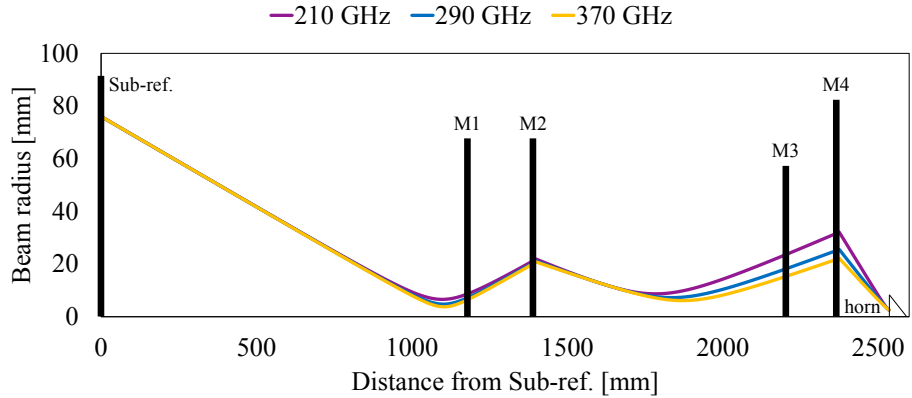} 
 \end{center}
\caption{Beam radius at 1/e of the electric field peak of the Gaussian beam as a function of the distance from the sub-reflector to the horn. At the sub-reflector and the horn aperture, the beam radii of 210, 290, and 370\,GHz are all equal. }\label{fig:optics_beam}
\end{figure}

It is required for the beam propagation system to couple 
the corrugated horn with the system without depending on the 
frequency.  By keeping the beam size and radius of curvature of the Gaussian beam constant at the sub-reflector and horn aperture independently on the frequencies, it is possible to maintain high aperture efficiency throughout the bandwidth.  We used the Frequency-Independent-Matching technique by \citet{Chu1983}, one of the design methods of frequency-independent optics. When designing, the beam size and radius of curvature at the sub-reflector and the diameter of the horn aperture were used as initial conditions. The distances between the sub-reflector and M2, between M2 and M4, and between M4 and horn were used as variables to find the optimal value of the radius of curvature at the horn aperture.  As a result, the beam radius at the horn aperture was calculated to be 2.57\,mm, and the radius of the curvature 32.5\,mm (\cite{Yamasaki2020}). When using a conical corrugated horn with a tapered structure, the coupling rate with the most basic Gaussian beam is the highest when the ratio of the radius of the Gaussian beam to the aperture radius in the aperture plane is 0.64, and the slant length is matched to the radius of curvature (see details in \cite{Goldsmith1998}). From these conditions, the aperture diameter of the corrugated horn was calculated to be 8\,mm. The semi-flare angle can be calculated from the horn diameter and the slant length (see figure \ref{fig:horn_pontie} for the definition of the parameters) and was estimated to be 7.07$^\circ$. Figure \ref{fig:optics_beam} shows the relationship between the distance from the sub-reflector and the beam sizes at 210, 290, and 370\,GHz. It is confirmed that the beam sizes at the sub-reflector and the horn aperture are the constant regardless of the frequency.

\section{Corrugated horn}
The aperture size and semi-flare angle of the corrugated horn were determined by the optics, as is shown in the previous section.  The remaining basic structural parameter is the inner diameter of the circular waveguide. It determines the cutoff frequency of the propagating electromagnetic wave mode, which greatly affects the frequency response. If the diameter is small, the cutoff frequency shifts to a higher value, causing degradation of characteristics, low return loss, on the lower frequency side. Note that the return loss is a measure of the effectiveness of power delivery from a transmission line to a load (for the definition, see \cite{Bird2009}). Therefore, the higher the return loss, the better the circuit is matched, i.e.,  the reflected power is lower. On the other hand, if the diameter is large, the cutoff frequency of higher-order modes shifts to a lower value, causing higher-order modes on the high-frequency side.  It is also necessary to match with the waveguide diplexer (Masui et al.\ \yearcite{Masui2020}, \yearcite{Masui2021}) connected immediately after the horn.  As a result of the optimization based on the above, the diameter of the circular waveguide was set to 0.91\,mm. The cutoff frequency of the basic mode TE$_{11}$ is thus 192\,GHz. \par
\begin{figure}[t]
 \begin{center}
  \includegraphics[width=12cm]{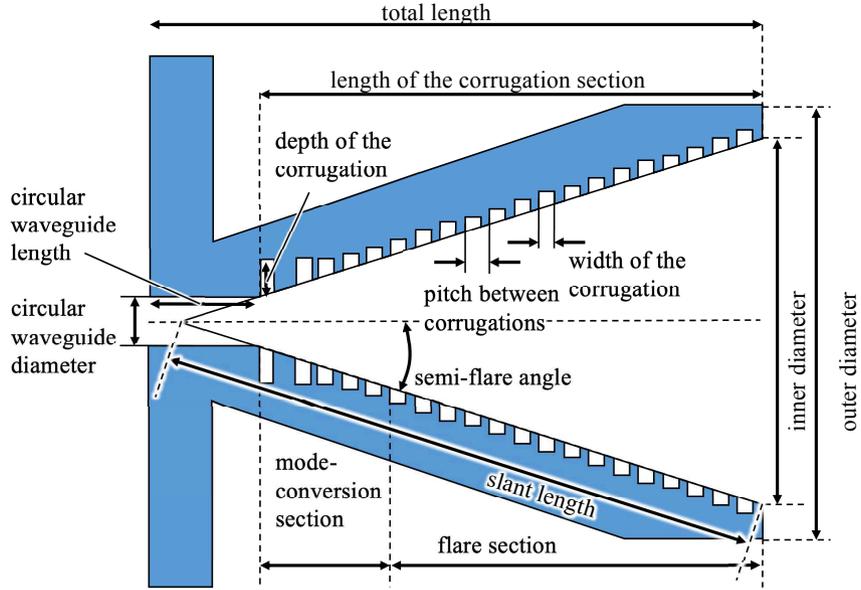} 
 \end{center}
\caption{Model drawing of the cross-section of a corrugated horn. The number of corrugations shown in this figure is different from the actual number. The whole structure is divided into a mode-conversion section and a flare section. The definition of each parameter is shown.} 
\label{fig:horn_pontie}
\end{figure}
\begin{figure}[h]
 \begin{center}
  \includegraphics[width=8cm]{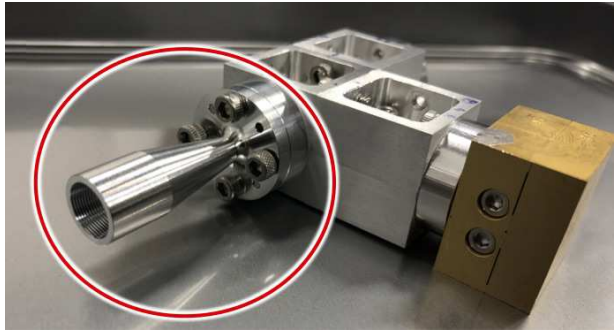} 
 \end{center}
\caption{Photograph of the corrugated horn attached to the multiplexer described in \citet{Masui2021}.}\label{fig:horn_picture}
\end{figure}

Next, the design procedure of the corrugation part is described. In the design of corrugated horns for radio telescopes, high return loss, low cross-polarization, and beam symmetry are required in order to efficiently irradiate the sub-reflector. The structure of the corrugations and their distribution determine the properties. Here, we set the goals of a return loss of 25\,dB or better and a maximum cross-polarization of $-$30\,dB or less at the simulation stage.  When the mode propagation of the corrugated horn is "balanced hybrid", the cross-polarized component is almost zero, and an axisymmetric and low cross-polarized beam is emitted. The condition for this is the presence of a sufficient number of corrugations that are narrow and have a depth of about a quarter of the wavelength (\cite{Clarr1984}). In this design, the choice of the depths of the corrugation is not straightforward because we will cover a wide frequency range; the optimum length of a quarter of the wavelength depends on the frequency. A depth of less than a quarter of a wavelength will result in an induced reactance, leading to the propagation of the EH$_{11}$ mode (\cite{Abbas-Azimi2009}), so the lowest frequency of 210\,GHz, corresponding to the wavelength of $\sim$1.4\,mm, is assumed in the depth calculation. There are about 70 corrugations in total, and it is not realistic to analyze all the corrugations simultaneously, including their depth, width, and pitch. The corrugated horn can be divided into two parts (figure \ref{fig:horn_pontie}); the mode-conversion section that converts the TE$_{11}$ mode in the circular waveguide to the HE$_{11}$ mode in the corrugations, and the flare section that determines the radiation characteristics of the horn. The first part is important for the matching bandwidth, whereas the second part is important to achieve the cross-polarization bandwidth. We optimized the mode-conversion section and the flare section separately in order to achieve the target performance over a fractional bandwidth of 56\%. A detailed analysis of each of these is discussed in the following sections. For these analyses, we used CHAMP (TICRA), a simulation software dedicated to corrugated horns and smooth-walled conical horns, which performs analysis based on the mode-matching method. With the software, we can calculate \textit{$S_{11}$} of the \textit{S}-parameter that is the negative of the return loss, and the beam pattern. The final horn parameters are summarized in table 2, and the photo is shown in figure \ref{fig:horn_picture}.

\begin{table}[t]
\label{tab:horn}
\tbl{Physical parameters of the corrugated horn}{
\begin{tabular*}{8.5cm}{ll}
\hline
Parameter & Value \\
\hline\hline
Total length                              & 31.44\,mm\\ 
Aperture outer diameter                   & 10.00\,mm\\
Aperture inner diameter                   & 8.00\,mm\\
Corrugation section length                & 28.44\,mm\\
Semi-flare angle                          & 7.07$^\circ$\\
Circular waveguide length                 & 3.00\,mm\\
Circular waveguide diameter               & 0.91\,mm\\
Corrugation pitch at flare section              & 0.40\,mm\\
Corrugation width at flare section            & 0.25\,mm\\
Corrugation pitch at mode-conversion section  & 0.17--0.46\,mm\\
Corrugation width at mode-conversion section & 0.13--0.27\,mm\\
\hline
\end{tabular*}
}
\end{table}

\subsection{Mode-conversion section}
\begin{figure}[h]
 \begin{center}
  \includegraphics[width=12cm]{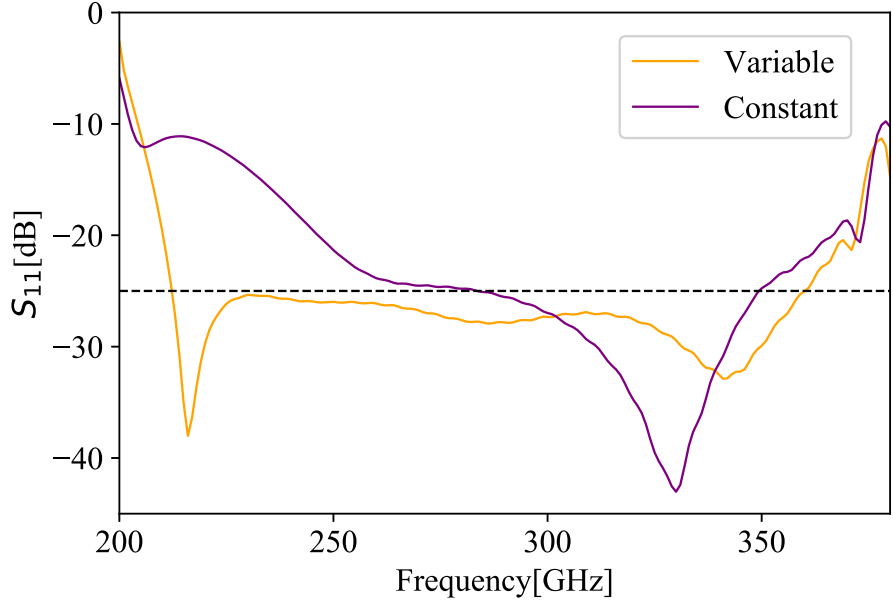} 
 \end{center}
\caption{The frequency response of \textit{$S_{11}$} when the width and pitch of the corrugation of the mode-conversion section are kept constant and when they are varied. In the constant case, the return loss is poorer especially at lower frequencies, but in the variable case, it is improved. The vertical axis is \textit{$S_{11}$} of an \textit{S}-parameter from the CHAMP simulation, which is the negative of the return loss.}
\label{fig:S11_pd}
\end{figure}

\begin{figure}[h]
 \begin{center}
  \includegraphics[width=12cm]{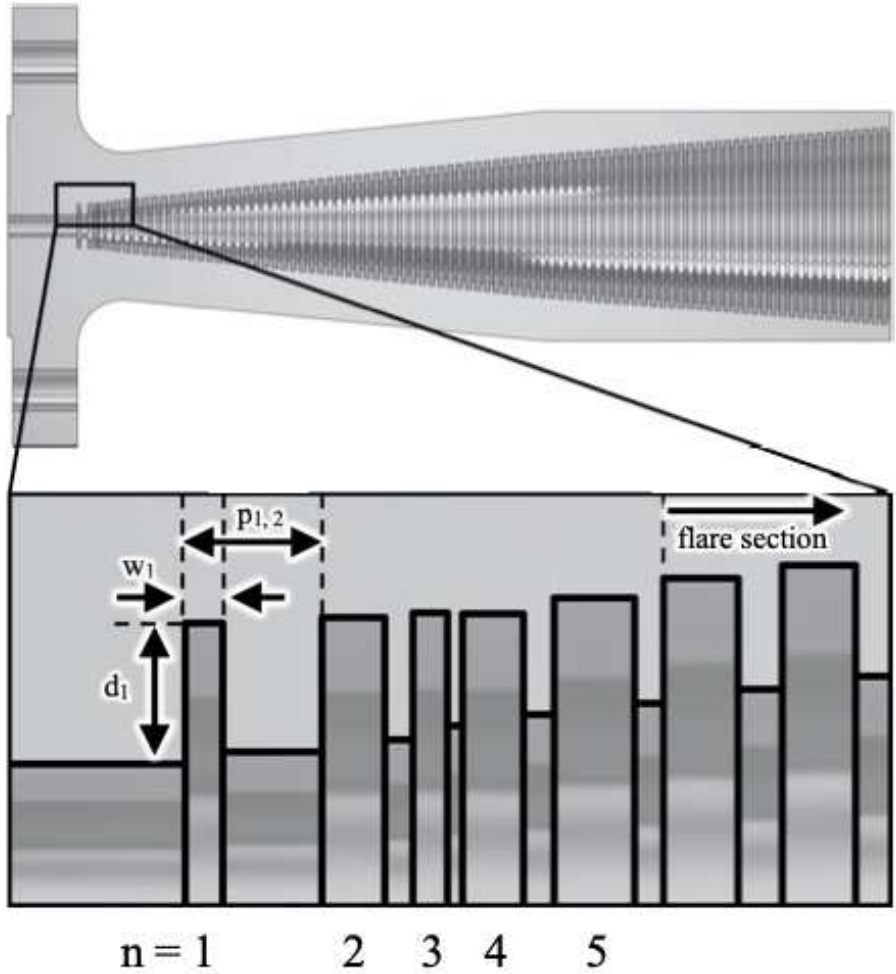} 
 \end{center}
\caption{The CAD drawing of the cross-section of the optimized mode-conversion section. It has variable pitches and widths. See table 3 for detailed corrugation parameters.} \label{fig:mode_conversion}
\end{figure}

The propagation of electromagnetic waves is reciprocal. We here consider the input of a radio wave from a circular waveguide to a horn. The input electric field mode is considered to be the fundamental mode TE$_{11}$ of the circular waveguide. The first five corrugations, which are the mode-conversion section, have a depth of $\lambda$/2--$\lambda$/4.  When TM$_{11}$ is excited in the section, and the ratio of TE$_{11}$ and TM$_{11}$ modes becomes 85\% and 15\%, the axisymmetric hybrid mode HE$_{11}$ is formed. The section also plays the role of impedance matching with the waveguide. The surface impedance of a corrugated waveguide with a corrugation depth of $\lambda$/4 is infinite, but that of a circular waveguide is zero, resulting in a large reflection. In the mode-conversion section, the first corrugation should have a depth of $\sim\lambda$/2, where the impedance is zero. Therefore, the depths of the mode-conversion section should change smoothly from $\lambda$/4 to $\lambda$/2 until reaching the flare section, which also affects how the mode-conversion occurs. The basic concept of the corrugated horn was described, for example, in \citet{James1981}, \citet{Dragone1977}. The mode-conversion section thus determines the overall performance of the corrugated horn. In particular, it was found to be decisive for the reflection characteristics, so we first optimized the return loss, i.e., impedance matching with the waveguide. When the pitch and width of the corrugations are constant, the maximum point of return loss is at 330\,GHz as shown in figure \ref{fig:S11_pd}. The return loss is evident at the lower frequency where the cutoff frequency is close. Therefore, we decided to change all the pitches, widths, and depths of the corrugations to optimize the performance. The depths, widths, and pitches of the mode-conversion section (number of corrugations: 5) were parametrically analyzed by CHAMP to find the optimum values while the depth of the flare section was kept constant.  We also set the constraint of the maximum aspect ratio, depth/width, of the corrugations.  If the aspect ratio is large, the cutting error will be large because a thin tool needs to be inserted deeper. Therefore, when optimizing the depth, we set the condition that the aspect ratio should be less than 4. The optimized values by the automatic parametric analysis are presented in figure \ref{fig:S11_pd}. The return loss is simulated to be higher than 25\,dB over the fractional bandwidth of $\sim$56\%. The detailed parameters of the optimized mode-conversion section (see figure \ref{fig:mode_conversion} for a clear definition) are shown in table 3.\par

\begin{table}[h]
\begin{center}
\label{tab:mode_converter}
\tbl{Parameters of the corrugations in the mode-conversion section. See figure \ref{fig:mode_conversion} for the definition.}{
\scalebox{1.3}{
\begin{tabular}{ccccccccc}
\hline
n     & \multicolumn{2}{c}{d$_{n}$ [mm]}&\multicolumn{2}{c}{w$_{n}$ [mm]}&\multicolumn{2}{c}{p$_{n, n+1}$ [mm]}\\ 
\hline \hline
1 & \multicolumn{2}{c}{0.481}& \multicolumn{2}{c}{0.129}&\multicolumn{2}{c}{0.465} \\
2 & \multicolumn{2}{c}{0.457}& \multicolumn{2}{c}{0.214}&\multicolumn{2}{c}{0.308} \\
3 & \multicolumn{2}{c}{0.432}& \multicolumn{2}{c}{0.119}&\multicolumn{2}{c}{0.170} \\
4 & \multicolumn{2}{c}{0.383}& \multicolumn{2}{c}{0.205}&\multicolumn{2}{c}{0.309} \\
5 & \multicolumn{2}{c}{0.396}& \multicolumn{2}{c}{0.275}&\multicolumn{2}{c}{0.374} \\
\hline
\end{tabular}
}
}
\end{center}
\end{table}

The outstanding feature of the optimized design is that the first corrugation is as narrow as 129\,{\textmu}m and is distant from the second corrugation. When the width and pitch were kept constant, and the depth is smoothly varied from $\lambda$/2 to $\lambda$/4, it was found that a narrower first corrugation resulted in a higher return loss at low frequencies by our simulation. This may be due to the fact that a narrower width results in better impedance matching with the circular waveguide, which improves the performance at low frequencies close to the cutoff frequency (see discussion in \cite{Zhang1993}). A simple formulation of the design of the corrugations after the second corrugations is difficult, but it seems that complex fine-tuning is performed in the simulation to keep the insertion loss low over the whole bandwidth.\par
The fabrication accuracy was measured to be $\pm$\,5\,{\textmu}m in the depth direction and $\pm$\,10\,{\textmu}m in the axial direction. We also performed a tolerance analysis to check whether cutting errors will have a significant impact on the performance of the design. In order to assess the impact on the performance due to the error, we first carried out a simulation by decreasing the depth of all the mode-conversion sections by 20\,{\textmu}m compared with the optimized design above as an example. In this case, the maximum return loss decrease is about 1\,dB around 225\,GHz. When the depth is 20\,{\textmu}m larger, the return loss falls short of 23\,dB in the frequency band above 350\,GHz. These results indicate that considering the current machining performance, the performance degradation due to the cutting error is small.


\subsection{Flare section}
\begin{figure}[h]
 \begin{center}
  \includegraphics[width=12cm]{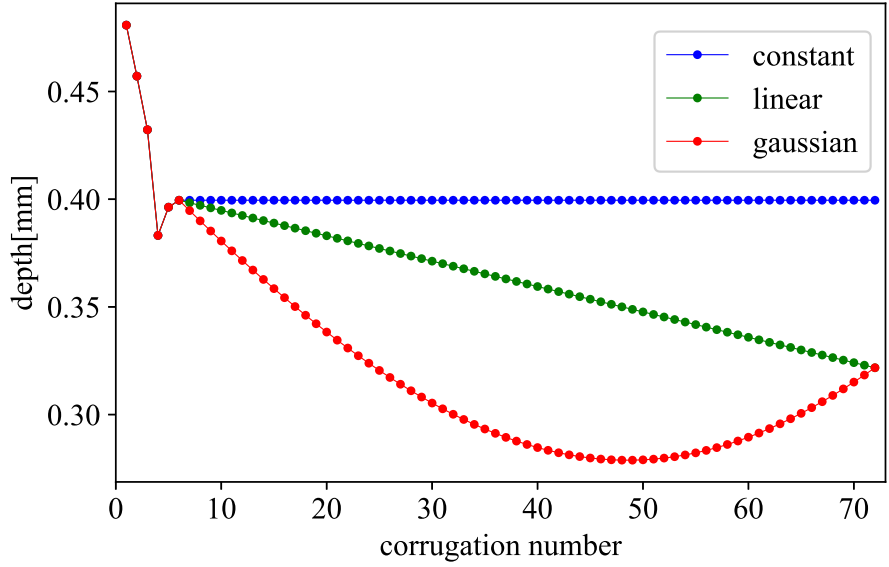} 
 \end{center}
\caption{Flare depth model used for optimization. The x-axis represents the number of corrugations counted from the waveguide side, and the first five are the mode-conversion section. The subsequent corrugation depths are shown for each of three models: 'constant',  'linear',  and 'gaussian'. }\label{fig:depth}
\end{figure}

After passing through the mode-conversion section, the signal enters the flare section. With the depth of the corrugation in the flare section fixed at the value of the last corrugation depth of the mode-conversion section, the maximum cross-polarization level was found to exceed $-$25\,dB at 350--360\,GHz.  In order to improve this cross-polarization characteristic, the following three models were developed and compared through simulations (figure \ref{fig:depth}).
\begin{enumerate}
  \item constant: Model with constant depth at the flare section.
  \item linear: Model in which the depth decreases linearly at the flare section.  
  \item gaussian: Model in which the depth of a corrugation is changed 
in proportion to the size of a Gaussian beam on a z-plane, assuming that a free space propagating inside a corrugated horn.

\end{enumerate}

\begin{figure}[h]
 \begin{center}
  \includegraphics[width=12cm]{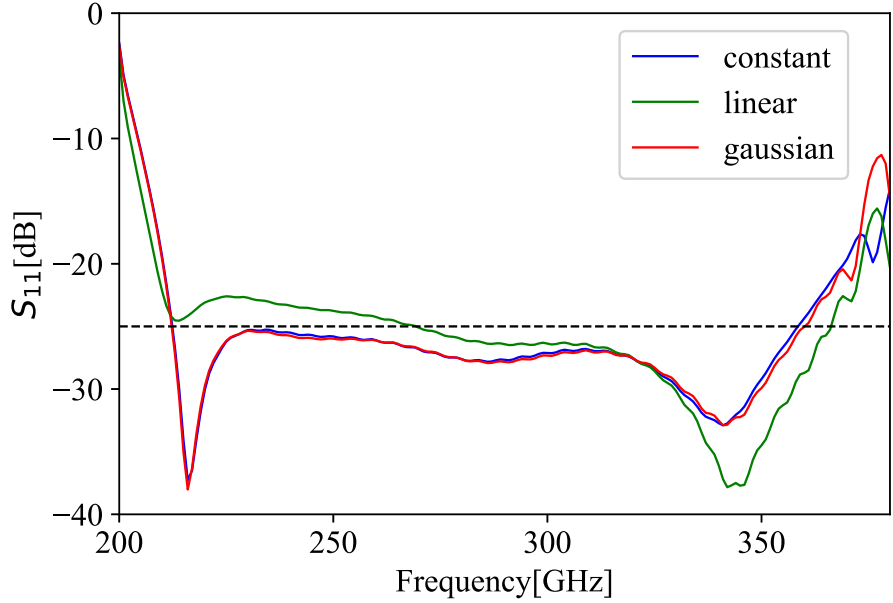} 
 \end{center}
\caption{Comparison results of \textit{$S_{11}$}, the negative of the return loss, in which the flare depth was simulated by 3 models; 'constant' model in blue, 'linear' in green, and 'gaussian' in red.} \label{fig:S11_comp}
\end{figure}

\begin{figure}[h]
 \begin{center}
  \includegraphics[width=12cm]{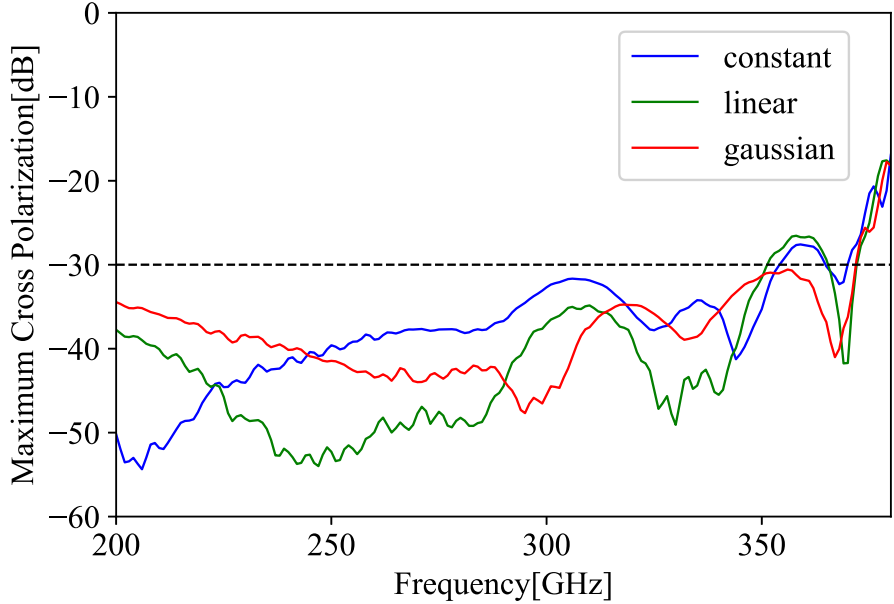} 
 \end{center}
\caption{The frequency response of the maximum cross-polarization for the simulated 3 models; 'constant' model in blue, 'linear' in green, and 'gaussian' in red. }\label{fig:Xs}
\end{figure}

'gaussian' is a model based on Gaussian Profiled Horn Antenna (GPHA). \citet{Salimi2013} optimized the GPHA model to follow and profile the propagation of Gaussian beams to achieve low sidelobes and maximum cross-polarization levels below $-$28\,dB at 10--17\,GHz. This idea is applied to the depth of the flare section. Assuming that the pitch and width of the flare are constant, three models are compared to achieve simulation results of higher than 25\,dB return loss and less than $-$30\,dB maximum cross-polarization.\par
The return loss is almost the same for 'constant' and 'gaussian', but slightly lower for 'linear' at low frequencies (20\,dB) (figure \ref{fig:S11_comp}). On the other hand, for cross-polarization, only 'gaussian' achieved the target value over the entire band (figure \ref{fig:Xs}). We thus adopted 'gaussian' for the depths of the flare section.

\section{Performances of the corrugated horn}

\subsection{Return loss}
We measured the \textit{S}-parameter using a vector network analyzer (PNA-X) at the National Institute of Information and Communications Technology (NICT). The measurement system consisted of an signal generator (SG) that outputs a 10--20\,GHz band and an extender that multiplies them. Then, a short-open-load-through (SOLT) calibration, which is known as a standard calibration method for an \textit{S}-parameter measurement system, of the two ports was performed. A black body inclined by 45$^{\circ}$ was placed at a distance of about 5\,cm from the horn aperture at the measurement. The measurement system is different for frequencies lower and higher than 330\,GHz. A WR-3.4 (0.864\,\textrm{$\times$}\,0.432\,mm) Extender and a waveguide transition from WR-3.4 to \textrm{$\phi$}0.91\,mm, a diameter of the circular waveguide of the corrugated horn, were used for the measurement in 220--330\,GHz. A WR-2.2 (0.570\,\textrm{$\times$}\,0.285\,mm) Extender and waveguide transitions from WR-2.3 (0.580\,\textrm{$\times$}\,0.290\,mm) to WR-2.8 (0.710\,\textrm{$\times$}\,0.356\,mm) and WR-2.8 to \textrm{$\phi$}0.91\,mm were used in 330--380\,GHz. Therefore, the result of the measured return loss is affected by the waveguide transitions, and is slightly different from the simulation result of the horn alone.\par

\begin{figure}[h]
 \begin{center}
  \includegraphics[width=12cm]{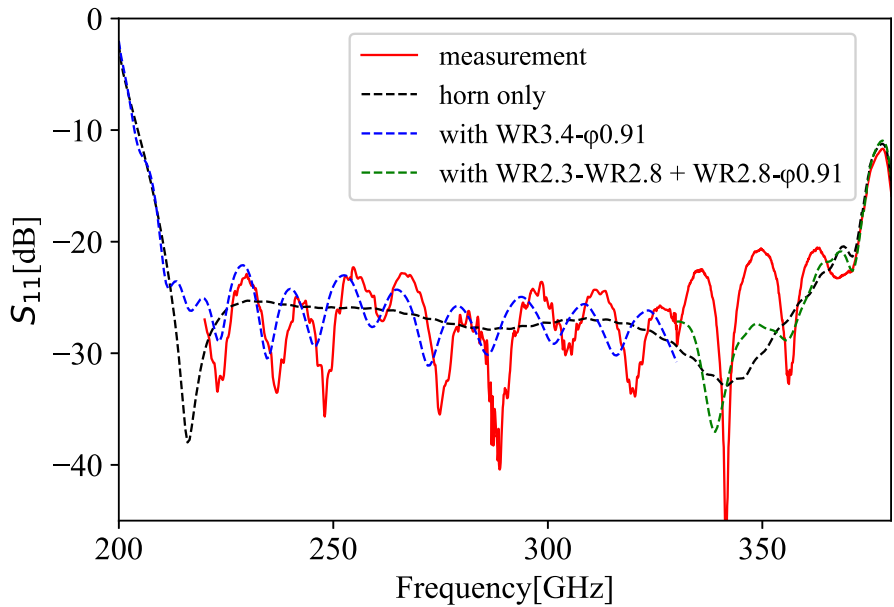} 
 \end{center}
\caption{The red line is the measured frequency response of \textit{$S_{11}$}, the negative of the return loss, and the broken lines are the simulation results by WASP-NET. Black shows the result without waveguide transitions, and blue and green with the waveguide transitions for 220--330\,GHz and 330--380\,GHz, respectively.} \label{fig:S11_WT}
\end{figure}

Here, we compare the measurement result with the simulated one.  The red line in figure \ref{fig:S11_WT} shows the measurement results. The result of the \textit{$S_{11}$} simulation of the corrugated horn is shown as the solid red line in figure \ref{fig:S11_comp}. Although the measurement \textit{$S_{11}$} roughly delineates the simulated one, there is a periodic response in the measured \textit{$S_{11}$}. This is because the waveguide transitions are attached to the corrugated horn, causing the standing-wave.  We thus used WASP-NET of Microwave Innovation Group for the simulation with the waveguide transitions included.  In figure \ref{fig:S11_WT}, the black broken line is the result of the simulation only for the corrugated horn, which is confirmed to be identical to the CHAMP simulation in figure \ref{fig:S11_comp}. The blue and green broken lines are that with the waveguide transitions, which roughly reproduces the standing-wave by the waveguide transitions. The measured result is thus consistent with the simulation, and the return loss is measured to be 20\,dB or better over the frequency range of 220--372\,GHz.

\subsection{Beam pattern of the horn}
\begin{figure}[h]
 \begin{center}
  \includegraphics[width=10cm]{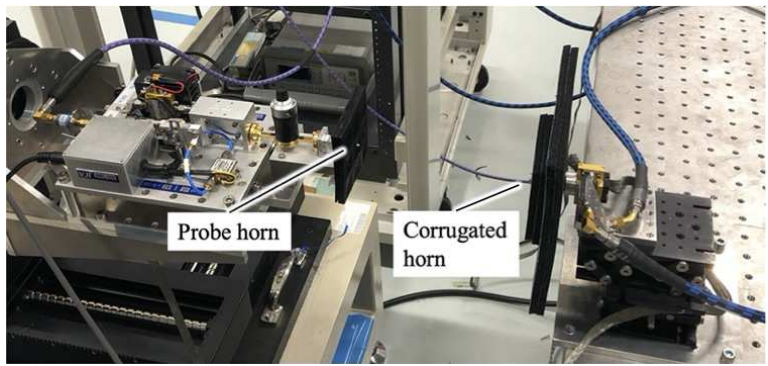} 
 \end{center}
\caption{Photograph of the near-field beam pattern measurement.}\label{fig:hm_pic}
\end{figure}

\begin{figure}[h]
 \begin{center}
  \includegraphics[width=10cm]{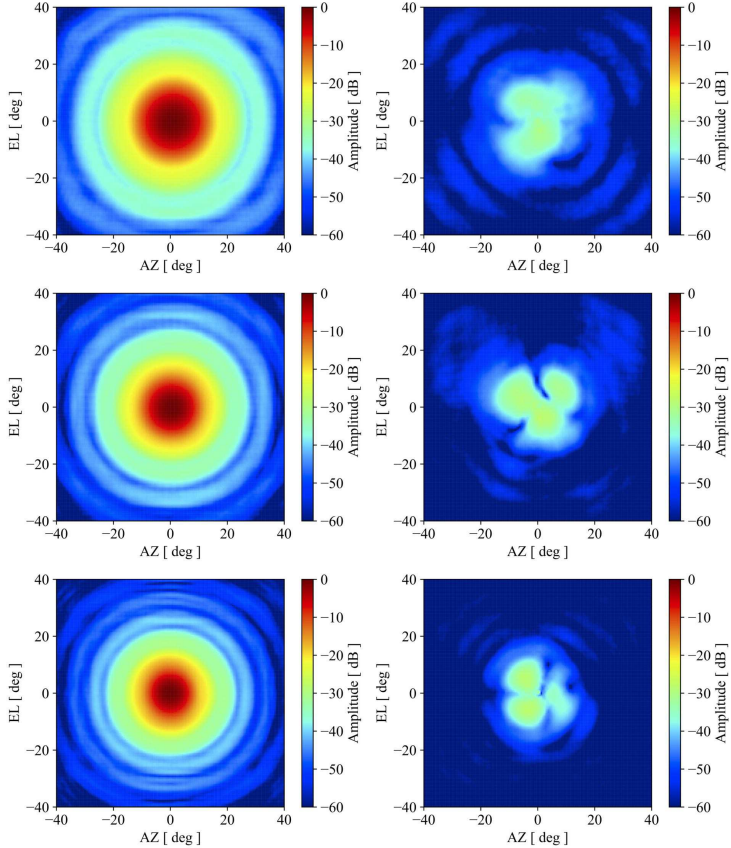} 
 \end{center}
\caption{Measured beam patterns of the corrugated horn in Az and El axes. From the top to bottom: the measurement results at the frequency of 230, 285, and 345\,GHz. The left is the pattern of the co-polarization, and the right the cross-polarization. }\label{fig:horn_grid}
\end{figure}

\begin{figure}[h]
 \begin{center}
  \includegraphics[width=10cm]{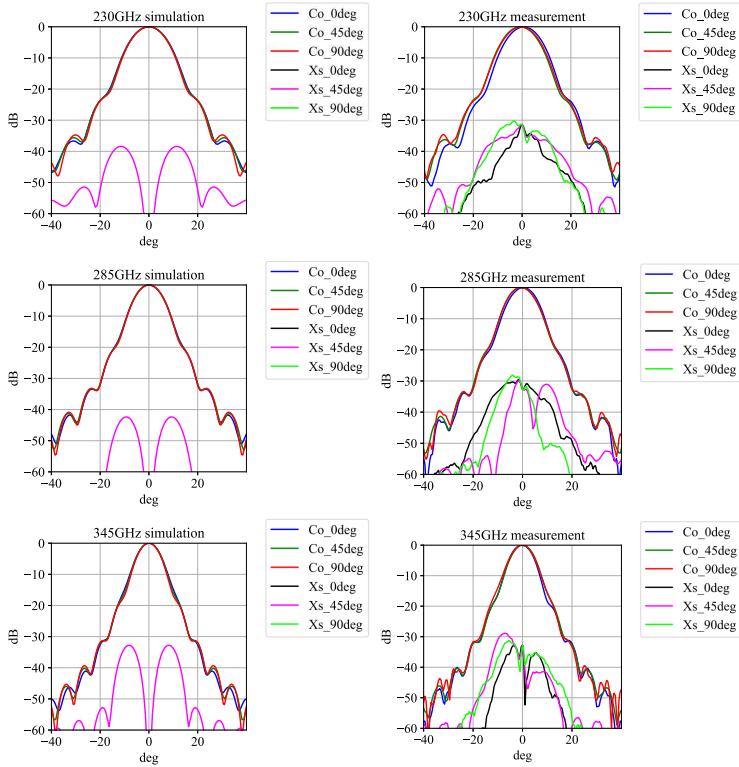} 
 \end{center}
\caption{Cuts patterns of figure \ref{fig:horn_grid} with three different cut angles (0, 45, and 90$^{\circ}$). From the top to bottom: the measurement results at the frequency of 230, 285, and 345\,GHz. The left is from the simulation, and the right is from the measurement.  The "Co" in the legend represents  "co-polarization", and "Xs" "cross-polarization", and the cut angle follows.}\label{fig:horn_cut}
\end{figure}

A reconfigurable beam measurement system (\cite{Gonzalez2016}) in the Advanced Technology Center of the National Astronomical Observatory of Japan was used to measure the beam patterns (figure \ref{fig:hm_pic}). The measurements were performed at three points: 230 and 345\,GHz, which are the emission line frequencies of the \textrm{$^{12}$}CO molecule, and 285 GHz in between. The beam patterns of the far-field are obtained by Fourier transformation of the measured beam patterns of the near-field. The beam maps in the Az, El plane are shown in figure \ref{fig:horn_grid}, and the cut patterns with three different cut angles (0, 45, and 90$^{\circ}$) are shown in figure \ref{fig:horn_cut}. Simulation results and measurements of the maximum cross-polarization level and HPBWs are summarized in table 4. The V-H difference of the HPBW is less than 0.2$^{\circ}$, which can be considered to be symmetric with respect to the axis and is in good agreement with the simulation. The cross-polarization levels are a bit higher than the simulation values; some are slightly above $-$30\,dB, although this is not a level that affects the observation performance.

\begin{table}[h]
\label{tab:horn_beam}
\tbl{Parameters of the far-field beam patterns of the corrugated horn. Maximum cross-polarization levels [dB] and half-power-beam-width [$^{\circ}$] in the V-H plane.}{
\scalebox{1}{
\begin{tabular}{lllllll}
\hline
\multicolumn{1}{l}{Frequency}               & \multicolumn{2}{c}{230\,GHz}                                          & \multicolumn{2}{c}{285\,GHz}                                          & \multicolumn{2}{c}{345\,GHz}                                          \\ \hline\hline
\multicolumn{1}{c}{}                                  & \multicolumn{1}{c}{Sim} & \multicolumn{1}{c}{Mes} & \multicolumn{1}{c}{Sim} & \multicolumn{1}{c}{Mes} & \multicolumn{1}{c}{Sim} & \multicolumn{1}{c}{Mes} \\ \cline{2-7}
\multicolumn{1}{l}{Max Xs.pol} & \multicolumn{1}{c}{$-$38}      & \multicolumn{1}{c}{$-$30}       & \multicolumn{1}{c}{$-$43}      & \multicolumn{1}{c}{$-$28}       & \multicolumn{1}{c}{$-$33}      & \multicolumn{1}{c}{$-$28}       \\
\multicolumn{1}{l}{HPBW (V)}         & \multicolumn{1}{c}{6.2}       & \multicolumn{1}{c}{6.0}        & \multicolumn{1}{c}{5.0}       & \multicolumn{1}{c}{5.0}        & \multicolumn{1}{c}{4.1}       & \multicolumn{1}{c}{4.1}        \\
\multicolumn{1}{l}{HPBW (H)}         & \multicolumn{1}{c}{6.3}       & \multicolumn{1}{c}{6.1}        & \multicolumn{1}{c}{5.1}       & \multicolumn{1}{c}{5.0}        & \multicolumn{1}{c}{4.3}       & \multicolumn{1}{c}{4.3}        \\ \hline
\end{tabular}}
}
\end{table}

\section{Evaluation of the corrugated horn and the optics on the 1.85-m telescope}

We installed the horn/optical system developed in this study on the 1.85-m telescope and realized the simultaneous observation of six molecular lines ($J$ = 2--1, 3--2) of CO isotopologues. The spectra toward Orion-KL and the OTF mapping results are shown in \citet{Masui2021}. Here, we derive the beam characteristics of the present system from the results of the commissioning observations from December 2020 to January 2021 with the remote operation.\par
The peak antenna temperature in \textrm{$^{12}$}CO($J$ = 2--1) toward Orion KL observed by the present system is $\sim$45\,K (\cite{Masui2021}), which is the same as \citet{Nishimura2015} with a fluctuation of $\sim$a few percent. The OTF mapping results in \textrm{$^{12}$}CO, \textrm{$^{13}$}CO, and C\textrm{$^{18}$}O($J$ = 2--1) are also consistent, including the temperature scale with the fluctuation.  For the observations toward IRC+10216, which is significantly compact for the beam size, the peak antenna temperatures are the same of about 1.2\,K within 10\% for both the previous and the present systems, which indicates that both beam size and main beam efficiency are virtually unchanged at 230\,GHz.\par
\begin{figure}[h]
 \begin{center}
  \includegraphics[width=12cm]{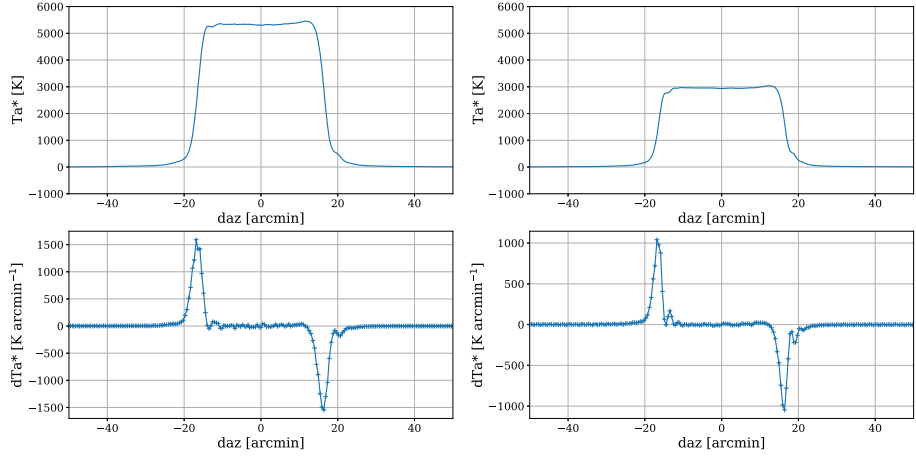} 
 \end{center}
\caption{The upper figures show azimuthal scans of the continuum level of the Sun. The vertical axis is the antenna temperature derived by the chopper-wheel method, and the horizontal axis is a displacement from the Sun center in arcminute. The output from Board 2, covering \textrm{$^{12}$}CO($J$ = 2--1), is in the upper left, and Board 4, covering \textrm{$^{12}$}CO($J$ = 3--2) in the upper right. The numbering of the Boards is from \citet{Masui2021}. The lower figures show the beam shapes derived by differentiating the upper results for Board 2 and Board 4.}\label{fig:sun}
\end{figure}

Next, we performed scan observations of the solar continuum level. By differentiating the intensity distribution at the edge of the Sun, it is possible to estimate the beam size (figure. \ref{fig:sun}). As a result of the measurement, the HPBW was measured to be 180$''$\,$\pm$\,6.$''$4 at 230\,GHz and 134$''$\,$\pm$\,7.$''$6 at 345\,GHz. It is to be noted that the saturation can occur at the SIS mixer or the IF chain because they are not designed for calibration with the Sun. Since the antenna temperature of the Sun observed in the 345\,GHz band is obviously lower than that in the 230\,GHz band (figure \ref{fig:sun}), the output in the 345\,GHz may have been largely saturated. We also note that the beam measurements by the planet scan could not be carried out due to the continuum level instability. The estimated 230\,GHz beam size from the solar scan is thus found to be consistent with the comparison with the previous CO observations described above. For 345\,GHz, since there are no previous 345\,GHz observations with the 1.85-m telescope that can be directly compared with the present observations, we check the beam efficiency and size at 345\,GHz by comparing the present results with those of other telescopes below.\par

\begin{figure}[h]
 \begin{center}
  \includegraphics[width=12cm]{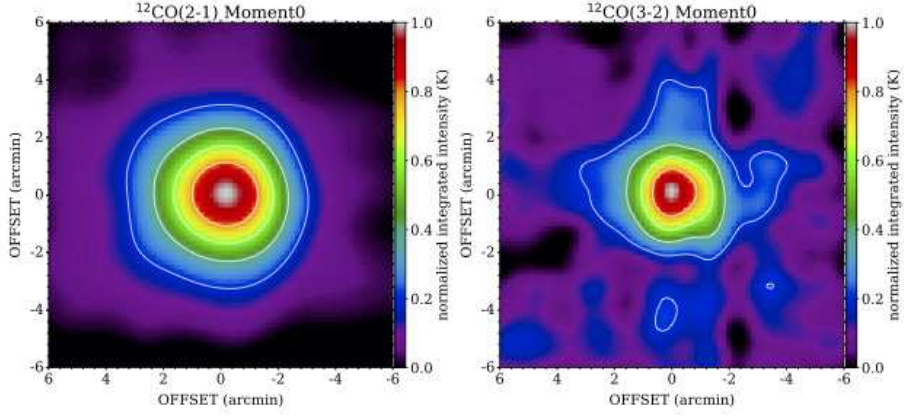} 
 \end{center}
\caption{CO({\sl J} = 2--1 and 3--2) simultaneous on-the-fly observation results of IRC+10216 observed using the receiver with the corrugated horn and optics we developed.}\label{fig:IRC}
\end{figure}

We compared our results with the \textrm{$^{12}$}CO($J$ = 3--2) observations toward IRC+10216 on a telescope with a relatively similar beam size (\cite{Tauber1989}) to check the beam size at 345 GHz. The beam size in their observations was 87$''$, and the source is considered to be sufficiently smaller than the beam size. The integrated intensity map of IRC+10216 obtained with the 1.85-m telescope is shown in figure \ref{fig:IRC}. The radial profile of the intensity of IRC+10216 can be assumed to be almost Gaussian, and its FWHM is 157$''$. From figure 4 of \citet{Tauber1989}, the FWHM in their observation is $\sim$120$''$. If we subtract the contribution of their deconvolved size of the target from our FWHM of 157$''$, we obtain our beam size of 134$''$, which is consistent with that from the solar scan. Assuming the main beam efficiency of the telescope to be 60\%, the peak antenna temperature expected from the model for an 87$''$ beam is $\sim$2.6\,K (figure 6 in \cite{Tauber1989}). Since the peak antenna temperature for sufficiently small targets is inversely proportional to the beam area, the peak antenna temperature for observations with a beam size of 134$''$ is estimated to be 1.1\,K. The peak antenna temperature of the present observation is measured to be $\sim$1.0\,K with an uncertainty of $\sim$10\%, which corresponds to the main beam efficiency of $\sim$55\% with an uncertainty of $\sim$10\%.\par

Next, to obtain the main beam efficiency at 345\,GHz independently, we compare it with the peak antenna temperature of \textrm{$^{12}$}CO($J$ = 3--2) toward Orion KL with the Mt. Fuji telescope. The Mt. Fuji telescope has a beam size of 132$''$ and a $T_{\mathrm{mb}}$  of 67.8\,K (\cite{Ikeda1999}). By smoothing our OTF mapping results to the Mt. Fuji telescope beam size, the peak antenna temperature is calculated to be $\sim$36.6\,K, and thus the main beam efficiency is estimated to be $\sim$54\%.\par
These results are summarized as follows; the beam size is $\sim$180$''$, and the main beam efficiency is $\sim$71\% in the 230\,GHz band, and the beam size $\sim$134$''$, and the main beam efficiency about $\sim$54\% in the 345\,GHz band. The designed values of the beam size at 230\,GHz and 345\,GHz are 172$''$ and 112$''$, respectively. The beam size at 230\,GHz is consistent within 5\%, while that at 345\,GHz is about 20\% larger.\par 
To investigate the cause of this large beam size at 345\,GHz, we simulated the effect of the displacement of the sub-reflector with respect to the main reflector. The beam size and beam efficiency are more affected by the displacement of the sub-reflector at higher frequencies. In the simulation, the displacement of the sub-reflector that makes the beam size 134$''$ at 345\,GHz is 0.7\,mm. Based on the aperture efficiencies with the displacement, the main beam efficiencies at 230\,GHz and 345\,GHz are calculated to be 84\% and 76\%, respectively, using the formula of \citet{Maddalena2010} and assuming $\eta_{\mathrm{R}}$ is 1. The degradation of the main beam efficiencies is caused by the surface roughness of the main reflector, misalignment of the optics, and the radome's transmission loss. Here, we assume all of these factors are caused by the mirror surface's roughness and determine the surface accuracy.
 Using the Ruze formula (\cite{Ruze1966}) and assuming a surface accuracy of 39\,{\textmu}m, the main beam efficiencies at 230\,GHz and 345\,GHz are calculated to be 73\% and 55\%, respectively, consistent with the present measurement results. The displacement of the sub-reflector this time seems to be due to the adjustment at 230\,GHz, where the atmospheric condition at 345\,GHz is far worse at Nobeyama Radio Observatory. If the sub-reflector can be set in the optimal location, the main beam efficiency is expected to be 69\% and the beam size 112$''$ at 345\,GHz. In summary, the measurement results on the beam are consistent with the designed performance when incorporating the sub-reflector displacement and the surface accuracy.

\section{Summary}
We developed a broadband conical corrugated horn and the corresponding optical system for the Osaka 1.85-m mm-submillimeter telescope covering the 210--375\,GHz band, and demonstrated the observations by using the system. The results of this paper can be summarized as follows:\par
1. By applying the Frequency-Independent Matching technique, a beam propagation system with high aperture efficiency independent of frequency was designed. The semi-flare angle, aperture, and waveguide size of the conical-corrugated horn were determined by the optics and subsequent components of the receiver.\par
2. A corrugated waveguide with a bandwidth of 210--375\,GHz (56\% fractional bandwidth) was designed. For a corrugated horn, it is necessary to carve corrugations sufficiently smaller than the wavelength, and since the wavelength is short in mm-submm, the flexibility in designing the corrugation is very limited. This makes the design of a wideband corrugated horn difficult in this frequency band. Therefore, we set a realistic limit on the corrugation design and focused on the following points. The return loss of the horn is improved by optimizing all the width, pitch, and depth of the corrugations, in the mode-conversion section consisting of the first five corrugations. The optimization was done by automated parametric analysis due to a large number of corrugation parameters. It was found that the narrower width of the first corrugation is important for smooth impedance matching, and the subsequent corrugation parameters need to be fine-tuned to have a better frequency response throughout the entire frequency range. The cross-polarization was reduced by changing the depths of the flare section with a gaussian shape.\par
3. The fabricated corrugated horn measurements indicate that the return loss is better than 20\,dB, and the cross-polarization lower than $-$25\,dB throughout the frequency range of 220--372\,GHz.\par 
4. The developed system was installed on the Osaka 1.85-m mm-submm telescope, and the simultaneous observations of \textrm{$^{12}$}CO, \textrm{$^{13}$}CO, and C\textrm{$^{18}$}O ($J$ = 2--1, 3--2) were successfully made. The beam size is $\sim$180$''$, and the main beam efficiency is $\sim$71\% in the 230\,GHz band, and the beam size $\sim$134$''$, and the main beam efficiency about $\sim$54\% in the 345\,GHz band. The measurement results on the beam are consistent with the designed performance when incorporating the sub-reflector displacement and the surface accuracy. If the sub-reflector can be set in the optimal location, the main beam efficiency is expected to be 69\% and the beam size 112$''$ at 345\,GHz.

\section*{Funding}
This work is supported by JSPS KAKENHI Grant Numbers JP18H05440, JP20J23670, JP15K05025, and JP26247026. This work was also supported by the grant of Joint Development Research supported by the Research Coordination Committee, National Astronomical Observatory of Japan (NINS).

\section*{ACKNOWLEDGMENTS} 
The authors are grateful to Issei Watanabe and Satoshi Ochiai of NICT for measuring \textit{S}-parameter of the corrugated horn. We are also grateful to Masanori Ishino of Kawashima Manufacturing Co., Ltd. (KMCO) for conducting the high-precision fabrication of the corrugated horn. We wish to thank the entire staff of the NRO, and all members of the 1.85-m telescope team to install and operate the receiver system.

\end{document}